\documentclass[11pt]{article}

\usepackage{amssymb}
\usepackage{subfig}
\usepackage{color}
\usepackage{epsfig,amssymb,amsfonts,amsmath,graphicx,dsfont,cite,xfrac}
\usepackage{authblk}
\definecolor{mygray}{gray}{0.5}

\usepackage{cite}
\usepackage[colorlinks=true,linkcolor=blue,citecolor=red]{hyperref}

\newcommand{\be}{\begin{equation}}
\newcommand{\ee}{\end{equation}}
\newcommand{\bea}{\begin{eqnarray}}
\newcommand{\eea}{\end{eqnarray}}

\title{Comments on ``Quasi-coherent states for the Hermite oscillator'' [J. Math. Phys. {\bf 59}, 062104 (2018)]}

\author[${}$]{Kevin Zelaya}
\author[${}$]{Oscar Rosas-Ortiz}

\affil[${}$]{\footnotesize Physics Department, Cinvestav, AP 14-740, 07000
M\'exico City, Mexico}

\date{}
\begin{document}

\maketitle

\begin{abstract}
The paper of \"Unal [J. Math. Phys. {\bf 59}, 062104 (2018)], though worthy of attention, contains a conclusion that is in error and may mislead the efforts to extend his results. The aim of the present note is twofold: we provide a correction to such a conclusion and then we emphasize some missing points that are necessary to clarify the content of the paper.
\end{abstract}


\vskip2ex
\noindent
The construction of a Gaussian wave packet for the quantum oscillator with time-dependent mass $M(\tau)>0$ has been discussed in a recent issue of this journal \cite{Una18}. Using the notation of \cite{Una18}, hereafter U--paper, the packet is written as
\begin{equation}
\psi(x,\tau)=N(\tau) \exp \left[ {-\frac{ \omega(\tau)}{2}x^2+b(\tau)x} \right].
\label{UN1a}
\end{equation}
This has been correctly assumed as a solution of the Schr\"odinger equation (with $\hbar=1$)
\be
i \frac{\partial \psi(x, \tau)}{\partial \tau} = \frac12 \left( - \frac{\partial^2}{M(\tau) \partial x^2} + M(\tau) x^2 \right) \psi (x,\tau).
\label{schro}
\ee
Then, the time-dependent coefficient $\omega(\tau)$ of the quadratic term in (\ref{UN1a}) is properly expressed as 
\be
i \omega = M \left( \frac{\dot u}{u} \right) = M \frac{d}{d \tau} \ln u,
\label{w}
\ee
where $u$ solves the classical equation of the damped oscillator with natural frequency $\omega_{osc}=1$ and time-dependent damping parameter $\gamma= \frac{\dot M}{M}$. Two additional key expressions are derived in U--paper to define the packet (\ref{UN1a}); they are respectively labeled as Eqs.~(9) and (12) in \cite{Una18}, and are written as follows ($z^*$ stands for complex conjugate of $z$):
\be
b(\tau) = \frac{b(0)}{u(\tau)}, \quad (\Delta x)^2= \frac{1}{\omega + \omega^*} = \frac{ \vert u(\tau) \vert^2 }{
 i M(\tau) W[u,u^*; \tau] } = \frac{ \vert u(\tau) \vert^2 }{ i W(0)}.
 \label{fake}
\ee
The expression on the right-hand side of (\ref{fake}) considers the fact that, if $u$ and $u^*$ are linearly independent, the Wronskian $W[u,u^*; \tau]$ is equal to the quotient $W(0)/M(\tau)$, with $W(0)$ a  constant (in principle arbitrary). The latter is correctly noticed and reported in U--paper, see paragraph below Eq.~(14), p.~3. 

After introducing the concrete forms of $N$, $\omega$ and $b$ in (\ref{UN1a}), the resulting packet $\psi(x, \tau)$ is expanded in terms of the Hermite  polynomials. The result leads to complex-valued expansion coefficients $C_n(\tau)$, the squared modulus of which are given by
\begin{equation}
\vert C_{n}(\tau)\vert^{2}=\frac{e^{-\lambda}}{n!}\lambda^{n}, \quad \mbox{with}  \quad \frac{\lambda(\tau)}{2} = \frac{\vert b\vert^2}{w+w^* }.
\label{UN2}
\end{equation}
The above expression corresponds to Eq.~(26) of U--paper, and is presented there as ``time-dependent Poisson distribution''.  

The ``time-dependence'' of $\vert C_n(\tau) \vert^2$, presumably  acquired through $\lambda(\tau)$, is highlighted in the U--paper as one of its main results (see e.g. lines 5-6 in the abstract, and lines 7-8 in the first paragraph of section IV, p.~6). However, although duly identified as a Poisson distribution, the probability weight $\vert C_n(\tau) \vert^2$ defined in (\ref{UN2}) is not time-dependent, which can be verified at the elementary level. For if we introduce (\ref{fake}) in (\ref{UN2}), the center $\lambda$ of the Poisson distribution is reduced as follows
\be
\lambda(\tau) = 2 \left\vert \frac{b(0)}{u(\tau)} \right\vert^2 (\Delta x)^2 =  \frac{ 2 \vert b(0) \vert^2}{i W(0)},
\ee
so that $\lambda(\tau) =\mbox{const}$. The straightforward calculation shows that, by necessity, the constant $W(0)$ is pure imaginary such that $W(0) = -iW_0$, with $W_0 >0$. Thus, what the author of \cite{Una18} failed to observe is that his own expressions give rise to the conventional (i.e., no time-dependent) Poisson distribution
\be
\vert C_{n}(\tau)\vert^{2}=\frac{e^{-\lambda_0}}{n!}\lambda_0^n, \quad \mbox{with}  \quad \frac{\lambda_0}{2} =  \frac{ \vert b(0) \vert^2}{W_0}.
\ee
Clearly, the affirmation that packets (\ref{UN1a}), together with (\ref{w}) and (\ref{fake}), represent superpositions of states associated to time-dependent Poisson distributions (\ref{UN2}) is in error, and it may mislead the efforts for extending the results of \cite{Una18}. Our comment is addressed to clarify such a point, in order to avoid confusion to the readers interested in the subject.

Nevertheless, we would like to emphasize that the above do not discount the time-dependence of the packet $\psi(x, \tau)$ by itself. Indeed, the elements of the expansion are found to be given by
\be
\Phi_n (x, \tau)= \frac{1}{\sqrt{2^n n!}} \left( \frac{\omega + \omega^*}{2\pi} \right)^{1/4} H_n \left( \sqrt{ \frac{\omega + \omega^*}{2} } x \right) \exp \left( -\frac{\omega}{2} x^2 \right),
\label{noH}
\ee
where the time-dependent functions $H_n(z(\tau) )$ satisfy the orthonormality condition of the Hermite polynomials. The above expression is quoted as Eq.~(24) in U--paper.

Additional comments are necessary to clarify some points that are missing in \cite{Una18}. 

\begin{itemize}
\item[i)] 
The packet (\ref{UN1a}) is of Gaussian profile whenever $\omega + \omega^* := 2 \mbox{Re} (\omega) >0$. That is, the real part of $\omega (\tau)$ is definite positive at any time. Otherwise, the packet $\psi(x,\tau)$ would be not normalizable. 

\item[ii)] 
As $\omega$ is proportional to the logarithmic derivative of $u(\tau)$, condition $\mbox{Re}(\omega) >0$ is fulfilled whenever $\mbox{Re} (u) \neq0$ and $\mbox{Im} (u) \neq 0$ at any time. In other words, $u$ is neither real nor pure imaginary. 

\item[iii)] 
The packets $\psi(x,\tau)$ that ``reduce to the coherent states of the harmonic oscillator when the effects of the damping are neglected'' are referred to as  ``quasi-coherent states'' in U--paper (see  p.~3, lines 2-3). However, no general conditions are given in U-paper for the packet (\ref{UN1a}) to satisfy such a property. By simple inspection one realizes that the time-dependent damping parameter is negligible whenever 
\be
\left\vert \frac{\dot M}{M} \right\vert  << 1.
\label{cond}
\ee 
In the simplest case ($M = \mbox{const}$), the Schr\"odinger equation (\ref{schro}) acquires the conventional form and $\vert u(\tau ) \vert^2$ may be set equal to 1. Then, $\mbox{Re}(\omega) = \omega_0 >0$ reduces (\ref{noH}) to the usual expression of the solutions associated with the harmonic oscillator, as this might be expected. However, in the most general case, condition (\ref{cond}) is not achievable for arbitrary forms of the function $M(\tau)>0$ at any time (see the next item).

\item[iv)]
In Section~III, it is given a mass $M(\tau) = \exp( \gamma_0 \tau^2/2)$, which produces the damping parameter $\gamma(\tau) = \gamma_0 \tau$. Then, it is argued that the constant $\gamma_0$ can take any value since the function $u(x)$ provided in Eq.~(28) of \cite{Una18} requires no square integrability (see paragraph below Eq.~(28), p.~5, of \cite{Una18}). However, if $\gamma_0$ is a positive number the mass $M(\tau)$ is unacceptable because it diverges as $\tau \rightarrow \infty$. A solution to this problem is either to take $\gamma_0 <0$ or to rewrite the mass in the form $M(\tau) = \exp( -\gamma_0 \tau^2/2)$, with $\gamma_0 >0$. In both cases, the mass is finite at any time and converges to zero as $\tau \rightarrow \infty$. The radius of convergence may be shortened by taking $\vert \gamma_0 \vert <<1$. Further approximations can be calculated by using the asymptotic expressions of the confluent hypergeometric functions in the related solutions. Nevertheless, the damping parameter is linear in $\tau$, so the condition for ``quasi-coherence'' cannot be satisfied at arbitrary time. To get an acceptable mass leading to the proper damping parameter one may consider, for instance, the exponential form $M(\tau) = \exp(-\gamma_0 \tau)$ with $0<\gamma_0 <1$.

\item[v)] 
The name ``quasi-stationary state functions'', coined in U--paper for the element series (\ref{noH}), requires justification. That is, although the $\Phi_n (x,\tau)$ are orthogonal at fixed $\tau$, it is missing the observable to what they belong as ``state functions''. In other words, it is necessary to identify the spectral problem to which time-dependent functions like (\ref{noH}) are solutions. Otherwise, the physical meaning of packets (\ref{UN1a}) is unclear. 

\end{itemize}

\noindent
To conclude this note we would like to emphasize that a more general form to construct Gaussian wave packets has been already reported in \cite{Cas13,Sch13}. There, the packet is written as
\be
\Psi_{WP} (x,t) = N(t) \exp\left\{ i \left[ y(t) \widetilde x^2 + \frac{1}{\hbar} \langle p \rangle \widetilde x +K(t)
\right] \right\},
\label{our1}
\ee
where $\widetilde x = x- \langle x \rangle$ and $\langle p \rangle = m \dot \eta$, with $\langle x \rangle \equiv \eta$ the mean value of position. The concrete form of the normalization factor $N(t)$ and the purely time-dependent phase $K(t)$ are not relevant at the moment. The time-dependent coefficient of the quadratic term is assumed complex, $y= y_R+iy_I$, and obeys the complex Riccati equation
\be
\left( \frac{2\hbar}{m} \dot y \right) + \left( \frac{2\hbar}{m}  y \right)^2 + w^2=0.
\label{our3}
\ee
In turn, the maximum of $\Psi_{WP}(x,t)$ is located at $x=\eta$ and follows a classical trajectory determined by the Newtonian equation
\be
\ddot \eta + w^2 (t) \eta =0.
\label{our2}
\ee
The straightforward calculation shows that (\ref{our3}) is solved by the function $\frac{2\hbar}{m}  y= \frac{\dot \alpha}{\alpha} + \frac{i}{\alpha^2}$ \cite{Cas13}, where $\alpha$ is a solution of the Ermakov equation
\be
\ddot \alpha + w^2(t) \alpha = \frac{1}{\alpha^3}.
\label{erma}
\ee
The Gaussian wave packet (\ref{our1}) includes the coherent states reported in \cite{Har82} for the harmonic oscillator with time-dependent frequency $w(t)$, and covers not only Hamiltonians with time-dependent (harmonic) potentials, but also time-dependent Hamiltonians that describe dissipative systems where the energy is no longer a constant of motion \cite{Cas13,Sch13,Cru15,Cru16} (further discussion on the subject can be found in \cite{Sch18}). The case discussed in the U--paper is also straightforwardly achievable. Recent applications of the approach dealing with the Gaussian wave packet $\Psi_{WP}(x,t)$ defined in (\ref{our1})-(\ref{erma}) include the propagation of Gaussian beams in parabolic media \cite{Cru17,Gre17}. The approach can be also implemented in the generation of nonclassical states of light in a Kerr medium \cite{Leo15}.

\vskip1ex
\noindent
{\bf Acknowledgments:} The support from CONACyT (project number A1-S-24569) is acknowledged.


\end{document}